\documentclass[a4paper]{article}

\usepackage{INTERSPEECH2021}
\usepackage{url}
\usepackage{graphicx} 
\usepackage{enumitem}
\usepackage{multirow} 
\usepackage{xcolor}
\usepackage{tipa}

\title{Manipulation of oral cancer speech using neural articulatory synthesis}

\name{Bence Mark Halpern$^{1,2,5}$, Teja Rebernik$^{2,4}$, Thomas Tienkamp$^{4}$, Rob van Son$^{1,2}$, Michiel van den Brekel$^{1,2}$, Martijn Wieling$^{4}$, Max Witjes$^{3,4}$, Odette Scharenborg$^5$}

\address{
 $^1$University of Amsterdam,
  $^2$Netherlands Cancer Institute, Amsterdam, 
  $^3$University Medical Center Groningen,
$^4$University of Groningen, Groningen,
  $^5$Delft University of Technology, Delft, The Netherlands
 }

\email{b.halpern@nki.nl}

\begin{document}

\maketitle
\begin{abstract}

We present an articulatory synthesis framework for the synthesis and manipulation of oral cancer speech for clinical decision making and alleviation of patient stress. Objective and subjective evaluations demonstrate that the framework has acceptable naturalness and is worth further investigation. A subsequent subjective vowel and consonant identification experiment showed that the articulatory synthesis system can manipulate the articulatory trajectories so that the synthesised speech reproduces problems present in the ground truth oral cancer speech.

\end{abstract}

\section{Introduction}
\label{sec:introduction}

Oral cancer is a type of cancer where a tumour is located inside the oral cavity, most typically on the tongue or floor of the mouth. Approximately $530,000$ people get diagnosed with this condition every year worldwide \cite{shield2017global}, including around $1000$ in the Netherlands \cite{cijfers_nl}. To treat oral cancer, (part of) the tissues surrounding the tumour are removed during surgery, which subsequently affects the patients' speech. There is large uncertainty regarding how the speech will be impacted by the surgery. This uncertainty causes significant distress to the patients, affecting their quality of life \cite{Epstein1999}. 

A speech synthesis system that could predict how a patient's voice would sound after surgery (post-operative speech), based on a surgical plan or a biomechanical model, could help clinicians and patients to make informed decisions about the surgery and alleviate the stress of the patients. Such a system would enable comparing multiple surgical plans and choosing the one that provides the best speech outcomes. Even if it is impossible a achieve a good speech outcome, patients could be counselled through the process using the synthesised samples.

Therefore, our long term aim is to build a pathological speech prediction system that can be connected to a biomechanical model incorporating the surgical planning \cite{kappert2019interactive}.
Previous studies have successfully shown that voice conversion systems can synthesise pathological speech of varying severity \cite{halpern2021objective, huang2021towards, illa21_ssw}. A limitation of these voice conversion systems is that they cannot be connected to biomechanical models, therefore, are unsuitable for our task.


Articulatory synthesis could provide a way to connect biomechanical models to speech prediction. Articulatory synthesis is a way to synthesise speech by either physical simulation of the speech production process \cite{fels2006artisynth}, or by using biosignals relevant to the articulation process, so-called data-driven approaches. Examples of the latter include ultrasound tongue imaging (UTI) \cite{csapo2017dnn, grosz2018f0}, magnetic resonance imaging (MRI) \cite{van2019cnn}, permanent magnetic articulography (PMA) \cite{gonzalez2017direct, gonzalez2016silent, gonzalez2014analysis} and electromagnetic articulography (EMA) \cite{taguchi2018articulatory, tobing2017articulatory, liu2018articulatory}. Among these techniques, EMA will be used as it has higher temporal resolution and spatial accuracy than the other techniques.

\begin{figure}
    \centering
    \includegraphics[width=\columnwidth]{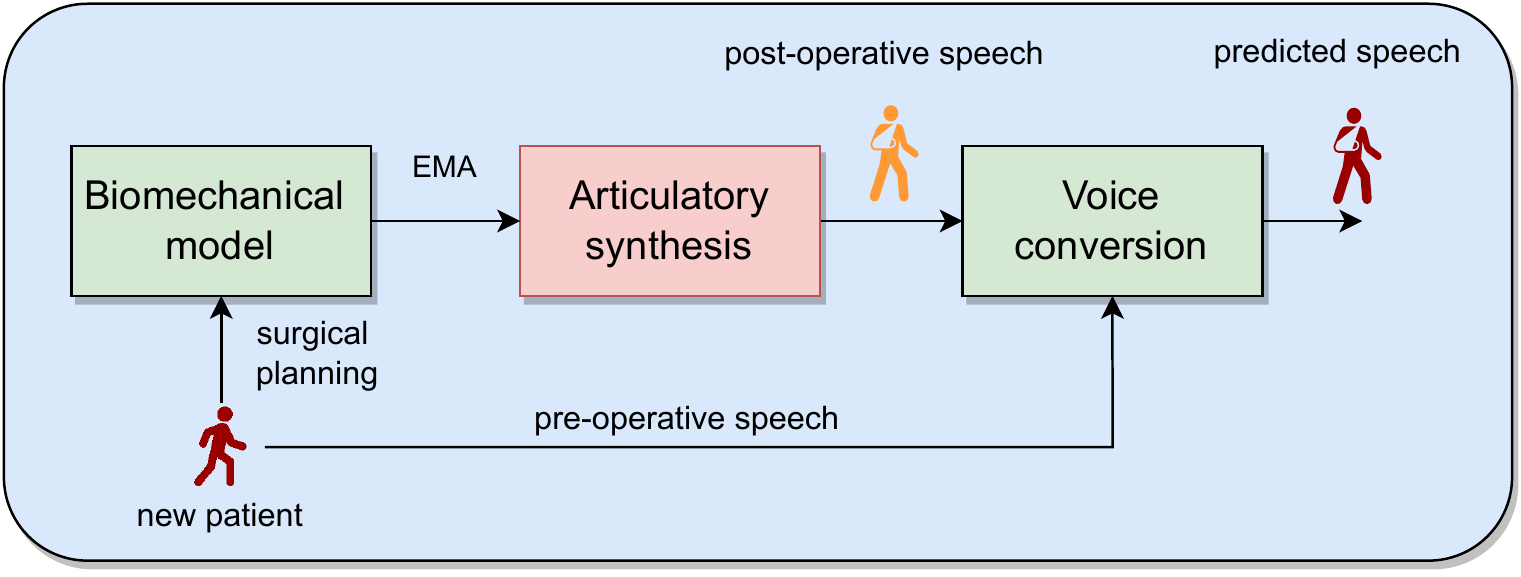}
    \caption{Outline of the current approach fitting into a future post-operative speech prediction framework. In the current study, we test the articulatory synthesis system (red box). The biomechanical model is presented in \cite{kappert2019interactive}, while the voice conversion is presented in \cite{illa21_ssw}. }
    \label{fig:outline_diagram}
    \vspace{-1.75\baselineskip}
\end{figure}

Figure~\ref{fig:outline_diagram} shows how the articulatory synthesis could connect the biomechanical model to the speech prediction. First, clinicians would plan the surgery with a new patient. The relevant surgical variables (i.e., which tissues to remove from the vocal tract) of this planning would then be incorporated in a biomechanical model. Then, tracking points in the biomechanical model could be converted into an EMA signal, which could be used to directly synthesise speech using the articulatory synthesis model. During training time, the articulatory system needs post-operative speech, which is impossible to obtain for the new patient (as this is the exact task for prediction). Therefore, at prediction time, the speech can be only synthesised with the vocal characteristics of a different (previously treated) oral cancer speaker. Thus, in order to achieve a patient-specific synthesis, the vocal characteristics of the synthesised speech need to be converted. This can be achieved using the voice conversion technique previously presented in \cite{illa21_ssw}, which uses the pre-operative speech and the synthesised post-operative speech to generate the new patient's post-operative speech. Finally, different post-operative samples could be synthesised using different surgical plans which could allow adjustment of surgical plans (which tissues/organs to spare), leading to better speech outcomes.

In the present paper, we carried out a feasibility study of such an oral cancer articulatory speech synthesis system (AS). We tested the AS in two different setups. The first setup (Synthesis) was a standard AS system, with the goal of synthesising new sentences based on articulatory trajectories, e.g., from a biomechanical model or EMA measurements. The second setup (Manipulation) tested whether it is possible to manipulate existing sentences by manipulating their articulatory trajectories slightly, which is often sufficient for clinical purposes. Specifically, we investigated two desired properties for the AS system: 1) the synthesised speech has to sound natural; 2) changes in the biomechanical model and the EMA trajectory should induce the correct acoustic changes.

\noindent
To summarise, we were interested in the following:
\vspace{-0.5em}
\begin{description}
\setlength\itemsep{0em}
    \item[RQ1] Is it possible to synthesise oral cancer speech using articulatory synthesis so that the synthesised speech has comparable naturalness to the (\textbf{a}) real (ground truth) oral cancer speech, (\textbf{b}) and to the synthetic (predicted) healthy speech?
    \item[RQ2] Can we perform phoneme-level manipulations on the articulatory trajectory in order to synthesise samples that correspond perceptually to the intended modifications?
\end{description}

The synthesised speech samples can be found online.\footnote{\url{http://slg.web.rug.nl/speech-samples/} password: interspeech2022}


\section{Dataset}
\label{sec:dataset}

To answer these questions, we collected a corpus consisting of 3.3 hrs of parallel speech and articulation recordings from seven native Dutch speakers. Five of the speakers had been diagnosed with and treated for oral cancer (three males and two females), the other two speakers were healthy controls (one male and one female).\footnote{We wanted to include more participants, however, the data collection of the research has been severely affected by the COVID-19.} The total duration of recordings per speaker varied between 24 min (nki01) and 32 min (nki03).
The speech was recorded in a sound-dampened booth using a Sennheiser ME66 microphone with a sampling frequency of 22,050 Hz. After the recording, the audio was downsampled to 16 kHz and mixed to mono. The speech and non-speech regions were then automatically annotated using Praat and then manually corrected \cite{Praat}.
The articulatory trajectories were recorded with the NDI-VOX electromagnetic articulograph, with a sampling frequency of 400 Hz, using 10 electrode channels. The recorded articulatory trajectories have a spatial accuracy of around 0.1 mm \cite{rebernik_accuracy}. Reference sensors (used to subtract head movements from articulator movements) were placed on both mastoids, the upper incisor, and the nasal bridge. Two movement sensors were placed on the mobile tongue, two on the vermillion border of the upper and lower lips, and one on the lower incisor or jaw. Not all movement sensors could be placed for every speaker.

The stimuli contained sentences from three sources. First, we selected  from the Wablieft newspaper corpus \cite{vandeghinste2019wablieft} sentences that together covered all Dutch phonemes and included many sentences with plosives, as these are known to be difficult for oral cancer speakers \cite{bressmann2004consonant, bressmann2009speech, Halpern2020}. Second, we included several Dutch texts that are commonly used for assessing speech impairment. The sentences in the first and second categories were produced once. Finally, we also included custom sentences into the stimuli to test the phoneme-level manipulation capability of the articulatory synthesis framework (\textbf{RQ2}). These custom sentences included five target words embedded in a carrier phrase and were repeated by the subjects five times in random order. 
The custom sentences share a similar form targeting (1.) vowels, (2.) and sibilants in CVC contexts:
\begin{enumerate}
\setlength\itemsep{0em}
\item \textbf{Hij heeft tamme \textit{baat}/\textit{biet}/\textit{boet} gezegd.} (\textipa{[bat]/[bit]/[but]})
\item \textbf{Hij heeft tamme \textit{sok}/\textit{shock} gezegd}.  (\textipa{[sOk]/[SOk]})
\end{enumerate}
There were a total of 226 sentences used as stimuli.
Table \ref{table:breakdown} shows the breakdown of the sentences. The full list of stimuli can be found online.\footnote{\url{https://karkirowle.github.io/groningen_corpus_perma/}}



\begin{table}[h]
\caption{The stimuli used in the collected corpus. The column Synthesis and Manipulation indicates the setup explained in Section~\ref{sec:experimental_design}.}
\vspace{-2 em}
\begin{center}
\resizebox{\columnwidth}{!}{
\begin{tabular}{c|c|c|c} 
 \toprule
 Name & Sentences & Synthesis & Manipulation \\ [0.5ex] 
 \midrule
   Wablieft \cite{vandeghinste2019wablieft} & 76 & Training & Training \\
 Papa en Marloes & 8 & Training & Training  \\ 
 Man uit Finland & 14 & Training & Training \\
 Noordenwind & 8 & Training & Training\\
 Els gaat naar markt & 10 & Training & Training \\
 Meneer van Dam & 6 & Training & Training \\ [1ex] 
 Jorinde en Joringel & 79 & Training [1-69] Test [70-79]& Training \\
 
 Custom & 25 & \multicolumn{2}{c}{See Section \ref{sec:experimental_design}} \\
 \midrule
 Total & 226 & 216 Train 10 Test & 211 Train 15 Test \\
 \bottomrule
\end{tabular}
}
\end{center}
\label{table:breakdown}
\vspace{-3 em}

\end{table}


\section{Design and methods}
\label{sec:design}

\subsection{Experimental design}
\label{sec:experimental_design}

All our experiments use the articulatory synthesis framework (AS), which is a speaker-dependent articulatory synthesis network (explained in Section~\ref{sec:architecture}). As Table \ref{table:breakdown} shows, the AS was trained with two different train-test partitioning of the dataset, which we named Synthesis and Manipulation. 


To test the capability of the AS on the synthesis task, the Synthesis setup contained no overlapping sentences between the training and the test set. We also created a different partitioning of the dataset that we called Manipulation. The Manipulation setup was based on including one variation of the custom sentences (and all the other sentences) in the training set and the other variations of the custom sentences in the test set: target words \textit{baat} and \textit{sok} were in the training set, while target words \textit{biet}, \textit{boet} and \textit{shock} were in the test set. The Manipulation setup allowed us to test explicitly whether the AS could synthesise phoneme-level differences. For example, a possible failure case would be that the AS copies the nearest example from the training set, i.e., even though \textit{boet} was said, \textit{baat} would be synthesised, as that is the closest example in the training set. We carried out subjective vowel and consonant identification experiments (see Section \ref{sec:vowel_consonant}) to test for this phoneme-level generalisation property (\textbf{RQ2}). Because the sentences in the test set are partially seen during training, we expected that this setup would also result in more natural speech. Note that such a setup is not unrealistic for our clinical scenario: using the Manipulation setup over the Synthesis setup has the considerable, but acceptable disadvantage that only sentences in the training set can be manipulated.

Objective naturalness evaluation experiments were performed on the test of the Synthesis and Manipulation, with additional subjective evaluation on the test set of Synthesis to answer \textbf{RQ1}, further explained in Section~\ref{sec:naturalness}.

\subsection{Speaker-dependent articulatory synthesis network}
\label{sec:preprocessing}

To achieve speaker-dependent articulatory synthesis, we used a neural network architecture based on the best practices from several prior studies on AS \cite{taguchi2018articulatory, cao2018articulation}. The neural network used the static, $\Delta$, $\Delta-\Delta$ articulatory trajectories as input features, which were downsampled to 200 Hz so that the sampling rate of the input and the output frames are matched. The neural network was a four layer LSTM with 128 units, and a final regression layer that matches the size of the output acoustic features. The output acoustic features were the static, $\Delta$, $\Delta-\Delta$ Mel-cepstrum (MCEP), which were extracted using the WORLD vocoder \cite{WORLD}.  

The predicted speech was then synthesised from the output acoustic features as follows. First, the output acoustic features were processed using maximum likelihood parameter generation (MLPG) to get a more robust MCEP estimation than the one provided by the neural network \cite{tokuda2000speech}. The $F_{0}$ and the BAP parameters required for synthesis were directly taken from the ground truth (copy synthesis) because we were primarily interested in the manipulation of the articulatory signals, and copy synthesis allowed more natural speech synthesis. Finally, the copy synthesis features were combined with the estimated MCEP features using the WORLD vocoder to obtain the final speech signal (predicted). We additionally synthesised samples using MCEP copy synthesis to measure the upper bound of naturalness (resynthesis).

In all cases, the neural network was trained with an Adam optimiser \cite{kingma2017adam}, using a learning rate of $0.001$, with early stopping until a maximum of $50$ epochs. Before training on the oral cancer data, a baseline was established using the \texttt{mngu0} dataset \cite{richmond2011announcing}, where our network obtained a Mel-cepstral distortion (MCD) of 6.4 dB (see Section~\ref{sec:naturalness} for more details on the MCD), which confirmed the correctness of our architecture implementation in a high-resource scenario. The source of the baseline is available online.\footnote{\url{https://github.com/karkirowle/articulatory_manipulation}}

Using the architecture described above, we trained a neural network from scratch on each speaker's data in both the Synthesis and the Manipulation setup. It was expected that the amount of training data would affect the results in both setups, therefore, data augmentation was performed by adding Gaussian noise with zero mean and four different standard deviation levels ($\sigma=10^{[0,-1,-2,-3]}$) to the articulatory trajectories. We repeated the data augmentation experiments with three different random seeds and performed a \textit{paired t-test} (paired by random seed) to check that the observed improvements are due to the data augmentation. 




\label{sec:architecture}

\subsection{Naturalness evaluation}
\label{sec:naturalness}

The naturalness of synthesised speech examples was evaluated with objective and subjective methods. To objectively evaluate the samples, we calculated the Mel-cepstral distortion (MCD) measure \cite{kubichek1993mel} on the entirety of the Synthesis and Manipulation test sets.

For subjective evaluation, we ran a five-point scale mean opinion score (MOS) based perceptual experiment, similar to the one carried out in \cite{huang2021towards}. Each listener rated 105 utterances: 5 random sentences from the Synthesis test set $\times$ 3 conditions (ground truth, resynthesis, predicted) $\times$ 7 speakers. We expect the naturalness to be lower in the pathological ground truth than in the healthy ground truth  based on our previous studies \cite{huang2021towards, illa21_ssw}. For all subjective evaluation experiments (including Section \ref{sec:vowel_consonant}), we used 9 native Dutch listeners. Note that the listeners were mostly from the Central Netherlands region, while the speakers were from the Northern region. 

\subsection{Vowel and consonant identification tests}
\label{sec:vowel_consonant}

We evaluated the success of the articulatory manipulation using a vowel and consonant identification test. In the vowel identification task, the listeners were given the sentence \textbf{Hij heeft tamme X gezegd} and asked to replace the X with \textit{biet}/\textit{baat}/\textit{boet}. Alternatively, listeners could indicate a different word via an input text field. 
In the consonant identification experiment, the listeners were provided with the sentence \textbf{Hij heeft tamme X gezegd} and had to choose from either \textit{shock}, \textit{sok} or write a different word via a text field as described above. 
Homonyms were not corrected.


In addition to the sentences predicted by the neural network, the stimuli for each experiment included the ground truth stimuli. Including ground truth is essential because we expect that the listeners might already have difficulties identifying vowels and consonants in the ground truth data. In total each listener rated 56 sentences for the consonant identification task (2 words $\times$ 2 conditions (ground truth, predicted) $\times$ 2 repetitions $\times$ 7 speakers), and 84 sentences (same, but 3 words) for the vowel identification task. Finally, to assess the impact of the manipulation statistically, a \textit{Fisher's exact test} was performed \cite{maclandrol}.


\section{Results and discussion}

\begin{table}
\caption{Objective (MCD in dB, lower is better) and subjective (MOS, higher is better) naturalness evaluation results. The columns contain the identifier and type of the speaker (healthy/oral cancer).  (\textdagger) indicates that the effect of data augmentation is statistically significant at $p < 0.05$. (*) in the case of the MOS indicates that the reduction in MOS is statistically significant at $p < 0.05$. }
\resizebox{\columnwidth}{!}{
\begin{tabular}{l|ccc|cccccc}
\toprule
 {} & \multicolumn{3}{|c|}{Healthy} & \multicolumn{6}{|c}{Oral cancer}  \\
 \midrule
{ID} & 01 & 07 & Avg & 02 &   03 &   04 & 05 & 06 & Avg \\
\midrule
Gender & male & female & & male & male & male & female & female & \\
\midrule
\multicolumn{10}{c}{SYNTHESIS} \\
\midrule
MCD & 7.91 & 8.67$^{\dagger}$ & 8.29 & 8.16 & 8.88$^{\dagger}$ & 7.16$^{\dagger}$ & 8.31$^{\dagger}$ & 8.48 $^{\dagger}$ & 8.19 \\
\midrule
MOS & \\
Ground Truth    &     4.73 &     4.89 & 4.81  & 3.96 &     4.02 &     3.37 &     4.33 &     4.17 & 3.96\\
Resynthesis&     4.13$^{*}$ &     4.73$^{*}$ & 4.43  &   3.4$^{*}$ &     3.39$^{*}$ &     3.24 &     3.93$^{*}$  &     4.03 & 3.61 \\
Predicted   &     2.28$^{*}$ &     2.87$^{*}$ & 2.57 &   2.19$^{*}$ &     1.93$^{*}$ &     2.23$^{*}$ &     2.62$^{*}$ &     2.26$^{*}$ & 2.24\\
\midrule
\multicolumn{10}{c}{MANIPULATION} \\
\midrule
MCD & 7.5$^{\dagger}$ & 7.36$^{\dagger}$ & 7.43 & 7.05$^{\dagger}$ & 8.06$^{*}$ & 6.58$^{*}$ & 7.79 & 7.3$^{\dagger}$ & 7.40\\ 
\bottomrule

\end{tabular}

\label{tab:mcd_result}
}
\vspace{-2 em}
\end{table}

\subsection{Naturalness results}


In the MCD rows of Table~\ref{tab:mcd_result}, we only report the results of the best data augmentation experiments, and it is also indicated whether the data augmentation improves the models (\textdagger) and if this improvement is significant (*). In the Synthesis case, data augmentation improved our models five out of seven times, but this improvement was never significant. In the Manipulation case, the data augmentation improved the models five out of seven times, and it was significant two times. Therefore, the noise-based data augmentation only mildly improved the naturalness of the predicted sentences compared to no augmentation.


From the MCD rows of Table \ref{tab:mcd_result}, we can see that the average objective naturalness scores of the oral cancer speakers are comparable to healthy speakers both in the Synthesis and Manipulation setup. Furthermore, it can be observed that the results are higher (worse) than we have obtained with our \texttt{mngu0} baseline (6.4 dB, see Section ~\ref{sec:architecture}). The only difference between our baseline and the AS models is the number of utterances: the \texttt{mngu0} training set contains 1226 utterances, while our dataset only contains 226. This observation suggests that the higher MCD results obtained are most likely due to the difference in the amount of data used.

From the ground truth MOS (subjective naturalness) rows of Table \ref{tab:mcd_result} we can see that the oral cancer speakers achieve lower MOS than the healthy speakers, which is in line with the results of our previous studies \cite{huang2021towards, illa21_ssw}. Furthermore, the resynthesis MOS row shows that the vocoder impacts the naturalness of the predicted speech: with the exception of nki04 and nki06 the vocoder makes the naturalness significantly worse (\textit{Wilcoxon}, $p<0.05$), which points towards the use of improved (neural) vocoders as a possible future improvement direction for the AS. Regarding the naturalness of the synthesised sentences we can see (predicted MOS row) that the synthesised results are significantly worse (\textit{Wilcoxon}, $p<0.05$) than the vocoder results, which means that there is room for improvement for the AS itself. The obtained MOS vary between 1.93 and 2.62. The mean MOS of the healthy speakers is slightly larger than that of the oral cancer speakers, but the difference is not statistically significant ($p=0.09$). Therefore, the AS has a similar performance in healthy and oral cancer speakers. 


We conclude that the naturalness of the predicted oral cancer stimuli is reasonable, although significantly lower than that of the ground truth oral cancer stimuli (\textbf{RQ1.a}). However, the naturalness of predicted oral cancer speech is comparable to the predicted healthy speech (\textbf{RQ1.b}). Slightly higher (poorer) MCD values are observed compared to our \texttt{mngu0} baseline, which is most likely due to the lack of training data. Furthermore, we observed lower (better) MCD values in the Manipulation setup than in the Synthesis setup, therefore Manipulation seems to be a more promising approach than Synthesis for our AS.


\vspace{-1 em}
\subsection{Vowel and consonant identification}

Neither vowel nor consonant identification mistakes with less than five replies ($n<5$) will be reported. Table~\ref{tab:consonant_confusion} (top) shows the results of the vowel identification experiment. No recognition errors are observed with the healthy control and healthy predicted speech that have more than 5 replies.




For the oral cancer ground truth speech, we can see that \textit{biet} \textipa{[bit]} is commonly misidentified. The most common misidentification pair is \textit{buut} \textipa{[byt]} (14.4\%), followed by \textit{bit} [bIt] (7.7\%).  In the case of predicted speech, we find that \textit{biet} \textipa{bit} (62.2\%, n=56) is often misidentified as \textit{buut} (25.6\%, n=23), \textit{bit} (10.0\%, n=9). Additionally, we find that \textit{boet} \textipa{[but]} (82.2\%, n=74) is misidentified as \textit{biet} (7.7\%, n=7), \textit{buut} (6.6\%, n=6).
Overall, the difference between predicted and ground truth vowels is not statistically significant ($p = 1$) for \textit{baat}, but significant for \textit{biet} ($p = 0.03$) and \textit{boet} ($p < 10^{-4}$). The misidentifications with \textit{biet} and \textit{boet} suggest that duration and tongue height differences need to be modelled better by the AS.

Table~\ref{tab:consonant_confusion} (bottom) shows the results of the consonant identification experiment. We can see that in the healthy case, we do not have any misidentification. For the predicted case, about half of the time \textit{shock} \textipa{[sOk]} is perceived as \textit{sok} \textipa{[SOk]}: upon inspection of the data, it turned out that these results are mostly exclusive to nki07, and therefore most likely a speaker-specific issue.

For the oral cancer ground truth speech, \textit{shock} is misidentified as \textit{sok} (10\%, n=10). \textit{Sok} was also identified as \textit{shock} (12.5\%, n=8). This \textit{shock}/\textit{sok} misidentification is in line with existing evidence, which finds that sibilants are impacted in oral cancer speech \cite{Halpern2020, jacobi2013acoustic}. In the case of the predicted samples \textit{shock} is misclassified as \textit{sok} (20\%, n=16) and \textit{sok} is misclassified as \textit{shock}   (13.8\%, n=11). Overall, the difference between predicted and ground truth are not statistically significant for \textit{shock} ($p=0.81$) and \textit{sok} ($p=0.39$). Therefore, the results suggest that sibilant manipulations are modelled well with our approach.

Overall, we find that the AS system has a promising ability to manipulate both healthy and oral cancer speech on the phoneme-level: we found no significant issues with the sibilants, but the front/back, height and vowel duration aspects have to be improved (\textbf{RQ2}).





\begin{table}[]
\caption{Confusion matrix showing the outcome of the vowel identification (top) and consonant (bottom) identification experiment.}
\vspace{-1 em}
\resizebox{\columnwidth}{!}{
\begin{tabular}{c|cccc|cccc}
\toprule
  & \multicolumn{8}{c}{Ground Truth}                                          \\
 \midrule
 & \multicolumn{4}{c}{Healthy}         & \multicolumn{4}{c}{Oral cancer}     \\
 \midrule
   & BAAT    & BIET    & BOET    & OTHER & BAAT    & BIET   & BOET    & OTHER  \\
 \midrule
  BAAT & 100.0\% & 0.0\%   & 0.0\%   & 0.0\% & 100.0\% & 0.0\%  & 0.0\%   & 0.0\%  \\
  BIET & 0.0\%   & 97.2\%  & 0.0\%   & 2.8\% & 0.0\%   & 77.8\% & 0.0\%   & 22.2\% \\
  BOET & 0.0\%   & 0.0\%   & 100\%   & 0.0\% & 0.0\%   & 0.0\%  & 100.0\% & 0.0\%  \\
  \midrule
       & \multicolumn{8}{c}{Predicted}                                             \\
 \midrule \
  BAAT & 100.0\% & 0.0\%   & 0.0\%   & 0.0\% & 100.0\% & 0.0\%  & 0.0\%   & 0.0\%  \\
  BIET & 0.0\%   & 100.0\% & 0.0\%   & 0.0\% & 0.0\%   & 62.2\% & 0.0\%   & 37.8\% \\
  BOET & 0.0\%   & 0.0\%   & 100.0\% & 0.0\% & 0.0\%   & 7.8\%  & 82.2\%  & 10.0\% \\
 \bottomrule
\end{tabular}
}
\newline

\resizebox{\columnwidth}{!}{
\begin{tabular}{c|ccc|ccc}
\toprule
      & \multicolumn{6}{c}{Ground Truth}                              \\
      \midrule 
      & \multicolumn{3}{c}{Healthy} & \multicolumn{3}{c}{Oral cancer} \\
      \midrule 
      & SOK      & SHOCK    & OTHER & SOK       & SHOCK & OTHER   \\
      \midrule 
SOK   & 100.0\%  & 0.0\%    & 0.0\% & 88.8\%    & 10.0\%    & 1.2\%   \\
SHOCK & 0.0\%    & 100.0\%  & 0.0\% & 12.5\%    & 85.0\%    & 2.5\%   \\
\midrule
      & \multicolumn{6}{c}{Predicted}                                 \\
      \midrule
SOK   & 96.9\%   & 3.1\%    & 0.0\% & 85.0\%    & 13.8\%    & 1.2\%   \\
SHOCK & 50.0\%   & 50.0\%   & 0.0\% & 20.0\%    & 76.2\%    & 3.8\%  \\
\bottomrule 
\end{tabular}
}
    \label{tab:consonant_confusion}

\vspace{-2 em}
\end{table}



\section{Conclusions}

In this paper, we presented an articulatory synthesis framework for the synthesis and manipulation of oral cancer speech for clinical decision making and alleviation of patient stress. Objective and subjective evaluations carried out on the articulatory synthesised speech demonstrated that the framework has reasonable naturalness on the synthesis of new sentences. The results are slightly poorer compared to our high-resource baseline most likely due to the lack of training data, which is only mildly alleviated by noise-based data augmentation. The system achieves good objective naturalness on the manipulated sentences. Our vowel and consonant identification experiment indicate that our articulatory synthesis system is able to manipulate sibilants and reproduce perceptual problems that are found in the ground truth oral cancer speech, such as the weak contrast between the sibilants \textit{sok} and \textit{shock}. Future work needs to investigate more advanced neural vocoders, improved sequence-to-sequence architectures to model vowel duration better, and better placement of electrodes to capture appropriate cues of tongue frontality and height. Speaker-independent articulatory synthesis could be also used to use to increase the amount of data available for training, and likely, increase the naturalness of the synthesised speech. Our results suggest that it is feasible to perform certain articulatory manipulations in data-driven systems, which is an important step towards enabling the connection of patient-specific biomechanical models to speech synthesis systems. 

\vspace{-1.5 em}
\section{Acknowledgements}

The authors would like to thank Finnian Kelly and Róbert Tóth for their constructive comments on the manuscript. This work received ethical clearance (NL76137.042.20). B.M.H. is funded through the EU’s H2020 research and innovation programme under MSC grant agreement No 766287. The Department of Head and Neck Oncology and Surgery of the NCI receives a research grant from Atos Medical (H\"orby, Sweden), which contributes to the existing infrastructure for quality of life research.

\bibliographystyle{IEEEtran}

\newpage
\bibliography{mybib}

\end{document}